\documentstyle[12pt]{article}
\if@twoside
 \else
 \fi
 
 \parskip \medskipamount

\begin{document}

\def\intt{\int \frac{d^3 k}{(2\pi)^3}}
\def\intl{\int \frac{d^3 l}{(2\pi)^3}}
\def\beq{\begin{equation}}
\def\eeq{\end{equation}}
\def\bqr{\begin{eqnarray}}
\def\eqr{\end{eqnarray}}
\def\ls{l \!\!\! /}
\def\ks{k \!\!\! /}
\def\qs{q \!\!\! /}
\def\ps{p \!\!\! /}
\def\ubar{\bar u}

\hfill{UMD-PP-95-32}
 
\vspace{24pt}
\begin{center}
{\large\sc{\bf Anomalous Magnetic Moment of Anyons}}
\baselineskip=12pt
\vspace{35pt}
 
Gil Gat$^{1}$ and Rashmi Ray$^2$
\vspace{24pt}
 
Physics Department\\
University of Maryland\\
College Park, MD 20742\\
U.S.A\\
\vspace{60pt}
\end{center}
\begin{abstract}
The anomalous magnetic moment of anyons is calculated to leading 
order in a $1/N$ expansion. It is shown that the gyromagnetic ratio g
remains 2 to the leading order in $1/N$. This result strongly supports 
that obtained in \cite{poly}, namely that g=2 is in fact exact.  
\end{abstract}
\vspace{86pt}
$^1$ggat@delphi.umd.edu
\newline
$^2$rray@delphi.umd.edu

\vfill
\baselineskip=24pt
\pagebreak

{\it Introduction}

In two spatial dimensions there is the intriguing possibility of anyons, 
particles with arbitrary spin and statistics \cite{wilc}. Although  there is
 currently no 
direct experimental evidence for the existence of anyons they may exist as
 quasi-
particles in realistic systems like the fractional quantum Hall system.

A consistent second quantized theory of anyons where the particle operators 
create and destroy anyons is yet to be formulated in a satisfactory way. In 
view of this, most analyses are performed with a theory of fermions (or bosons)
 coupled 
minimally to a gauge field whose action  contains a Chern-Simons (C-S) term.
 The C-S term, which provides topological mass for the gauge field , transmutes
 the fermion(or boson) spin and statistics to those appropriate for
anyons.

 A matter of vital importance is the investigation
of the electromagnetic interaction of anyons, in particular of the gyromagnetic
ratio (g). It has been demonstrated in \cite{kogan} that starting from a fermionic
 (bosonic)
theory the anomalous magnetic moment of anyons (at the one  loop level) is proportional to 
the induced fractional spin ,with a gyromagnetic ratio of 2.
Subsequent work \cite{poly} tends to suggest that this ratio remains 2 to all orders 
i.e. it is exact.

It is well known that the fractional spin is unrenormalized beyond one loop 
(Coleman-Hill theorem \cite{cole}). Thus for the g-factor to remain 2 to all orders in 
perturbation theory the anomalous magnetic moment ought to be unrenormalized
beyond one loop too. One might try to
extract this information from an analysis of the entire weak coupling 
expansion.
 It is not clear to us however what symmetry
if any will facilitate this task 
. Instead we have espoused in this work a more modest
 approach , namely the 1/N expansion. The 1/N expansion effectively resums 
an infinite subclass of diagrams (i.e. the bubble diagrams). It thus contains
information beyond weak coupling expansion and at least in the infinite
N limit it becomes exact. Except for lattice simulations, this method is
possibly the only systematic non-perturbative method one can use.

{\it The Model}

Consider a model with $2N+1$ identical fermionic species coupled minimally to
a gauge field in 2+1 dimensions
\begin{equation}
{\cal L}_E=\bar \psi_a (i \partial \!\!\! / +m-\frac{e}{\sqrt{N}}A \!\!\! /)
\psi_a-\frac{1}{4} F_{\mu \nu}
F^{\mu \nu}-i \frac{\kappa}{8\pi} \epsilon^{\mu \nu \rho} A_{\mu} 
\partial_{\nu} A_{\rho}.  
\end{equation}
An odd number of fermion flavors is necessary to satisfy the requirements
of P and T violation. This choice of lagrangian is quite general
since  as we shall see both 
a C-S term and a Maxwell term would have been generated radiatively even 
if our original lagrangian did not contain them.
 
In order to obtain a sensible expansion in powers of $1/N$, we fix
$e^2N=\gamma $ and reexpress the Lagrangian in eq.(1) in terms of
$\gamma $. The generating functional for the Green functions of the
theory is  
\begin{equation}
{\cal Z}=\int D\bar \psi D\psi DA \delta(\partial \cdot A) e^{-S_A-S_f}
\end{equation}
where 
\begin{eqnarray}
S_A &=&-\int d^3x [\frac{1}{4} F_{\mu \nu} F^{\mu \nu}+i \frac{\kappa}{8\pi} 
\epsilon^{\mu \nu \rho} A_{\mu} \partial_{\nu} A_{\rho}] \\ \nonumber
S_f &=& \int d^3x \bar \psi_a (i \partial \!\!\! / +m-{{\gamma
}\over{\sqrt{N}}}A \!\!\! /)\psi_a
\end{eqnarray}
Integrating over the fermionic fields we get
\begin{eqnarray}
{\cal Z}&=&\int DA \delta(\partial \cdot A) e^{-S_A-W_f}
\\ \nonumber
W_f &=& NTr\log(i \partial \!\!\! / +m)+NTr\sum_{n=1}^{\infty}
\frac{1}{n} (\frac{1}{\sqrt{N}})^n (\frac{1}{i \partial \!\!\! / +m} A\!\!\!
 /)^n
\end{eqnarray}
The lowest order term in $W_f$, after discarding the field independent term, 
vanishes on using any soft cut-off procedure that respects gauge invariance 
(e.g. dimensional regularization).The quadratic term in the field gives the 
leading order improved inverse propagator for the gauge field
inverse propagator for the gauge field (in the Landau gauge)
\begin{equation}
D(p)_{\mu \nu}=D(p)_{\mu \nu}^{tree}-\Pi(p)_{\mu \nu} \label{prop}
\end{equation}
where $D(p)^{tree}$ is given by $D(p)^{tree}=-p^2[ {\cal P}_{\mu \nu}+i
\frac{\kappa}{4\pi}\epsilon_{\mu \nu \rho} \frac{p^{\rho}}{p^2}]$  and   
\begin{equation}
\Pi_{\mu \nu}= i \int \frac{d^3k}{(2\pi)^3} \,  \frac{tr(\gamma_{\mu} (k 
\!\!\! / +m) \gamma_{\nu}
(k \!\!\! /-p \!\!\! /+m))}{[(k-p)^2-m^2][k^2-m^2]} 
\end{equation}
the 2+1 dimensional Dirac matrices satisfy 
\begin{equation}
\gamma_{\mu} \gamma_{\nu}=g_{\mu \nu}-i \epsilon_{\mu \nu \rho }\gamma^{\rho}
 \hspace{2cm}
tr(\gamma_{\mu} \gamma_{\nu})=2g_{\mu \nu} . \label{gid}
\end{equation}
Using dimensional regularization and standard Feynman parameters we find
\begin{equation}
\Pi_{\mu \nu}(p) = F(p) {\cal P}_{\mu \nu}+im \epsilon_{\mu \nu \rho} 
\frac{p^{\rho}}{p^2}G(p)  \label{prop1}
\end{equation}
where 
\begin{eqnarray}
{\cal P}_{\mu \nu} &=& g_{\mu \nu}-\frac{p_{\mu} p_{\nu}}{p^2} \\ \nonumber
G(p) &=& -\frac{p}{4\pi}\log \left(\frac{p+2m}{p-2m} \right)-\frac{ip}{4}
 \\ \nonumber
F(p) &=& \frac{1}{4\pi} \left[ m-\frac{p^2+4m^2}{4p}\log 
\left(\frac{p+2m}{p-2m}\right)-\frac{i \pi (p^2+4m^2)}{4p} \right]
\end{eqnarray}
Combining eq.(\ref{prop}) and (\ref{prop1}) for $D(p)_{\mu \nu}$ and 
inverting we get the leading order gauge propagator
\begin{equation}
G(p)_{\mu \nu}=-i\frac{(-p^2)\Pi_e}{\Pi_e^2 p^2-\Pi_o^2 M^2}\left[ 
{\cal P}_{\mu \nu}
-iM\frac{\Pi_o}{\Pi_e} \epsilon_{\mu \nu \rho} \frac{p^{\rho}}{p^2} \right]
\end{equation} 
where $\Pi_e=p^2+F(p)$ and $\Pi_o=p^2+G(p)$ are the even and odd parts of the 
inverse propagator \cite{jacktemp}.

 For large enough $M=\frac{\kappa}{4\pi}$ i.e. the 
anyons spin $S=\frac{1}{2}+\frac{1}{\kappa}$ is very close to $1/2$. Thus as 
has been discussed in \cite{shizuya}, an anyon may be viewed as a point charge
 surrounded by a gauge field cloud of size $\sim {1\over{M}}$. Thus in the 
limit $ M \gg m $,
the cloud collapses on to the charge and such a system of charge-cloud 
composites will display fractional statistics upon interchange even for 
arbitrarily small separation between the particles\footnote{ Actually for 
large enough $M$  $\Pi_e^2 p^2-\Pi_o M^2$ develops an
 imaginary part , so that the gauge field becomes unstable \cite{brown}, however the width of this resonance is very small and goes as $1/M$}
Thus such a limit as envisioned above seems to be indicated for a system of
 anyons.

{\it Anomalous Magnetic Moment}

The anomalous magnetic moment of fermions is defined by using the Gordon 
decomposition of the fermionic vector current. In 2+1 dimensions it has 
the following form:
\begin{equation}
\bar \psi(p+q) \gamma_{\mu} \psi (p)=\frac{(2p+q)_{\mu}}{2m} \bar \psi (p+q)
\psi (p) +i \epsilon_{\mu \nu \lambda} \frac{q^{\lambda}}{2m} \bar \psi (p+q)
\gamma^{\nu} \psi (p).
\end{equation}
By coupling this current to an external electromagnetic field  it is easy to 
see that the magnetic moment is the 
coefficient of $i \epsilon_{\mu \nu \lambda} \frac{p^{\nu} q^{\lambda}}{m}$
. The magnetic moment $\mu$ can be written as $\mu=\frac{g}{2m}S=\frac{1}{2m}$.

For fermions the spin $S=1/2$ so the gyromagnetic ratio $g=2$. For anyons 
$S=\frac{1}{2}+\frac{1}{\kappa}$ 
does not shift from its value which is determined at one loop \cite{kogan}.
 It was shown that to this
 order $g$ is the same as for fermions i.e. $g=2$.  Therefore if $\mu$ gets
 any correction it must alter the gyromagnetic ratio.
 In what follows we give the details of a calculation
of $\mu$ to leading order in $1/N$. This calculation goes beyond one loop
 because the leading order of $1/N$ already sums an infinite subset of weak 
coupling diagrams. In fig.1  we show the leading order diagrams that 
contribute to the anomalous magnetic moment. In addition to the usual weak
 coupling diagram ($a$) we have
a two loop diagram ($b$) that is of the same order
 $(\frac{1}{\sqrt{N}})^{3/2}$, however this diagram vanishes identically due 
to $C$ invariance. We thus consider only diagrams  ($a$). 

\begin{equation}
\Gamma_a(p,q)_{\mu}=-i \int \frac{d^3k}{(2 \pi^3)} \frac{\bar u(p-q)
 \gamma_{\nu} (p \!\!\! /
-q \!\!\! / -k \!\!\! +m) \gamma_{\mu} (p \!\!\! / -k \!\!\! / +m) \gamma_{
\lambda} u(p)}{[(p-k)^2-m^2][(p-k-q)^2-m^2]}G(k)^{\nu \lambda}
\end{equation}

The large $N$ gauge field propagator in eq. (\ref{prop}) has a very 
complicated form. In
 particular this form does not render itself for Feynman parametrization. It
 is therefore useful to use the K\"{a}llen- Lehmann representation for this 
propagator.
\begin{equation}
G(p)_{\mu \nu}=-i \left[ \int_0^{\infty} ds \frac{\rho_1(s)}{p^2-s^2+i 
\epsilon} 
{\cal P}_{\mu \nu} (p)-iM\epsilon_{\mu \nu \rho} \frac{p^{\rho}}{p^2} 
\int_0^{\infty}
ds \frac{\rho_2(s)}{p^2-s^2+i\epsilon} \right]
\end{equation}
where 
\begin{equation}
\rho_1(s)=\frac{s}{\pi} Im \frac{\Pi_e(s)}{\Pi_e(s)^2 s-\Pi_o(s)^2 M^2} 
\hspace{2cm} 
\rho_2(s)=\frac{s}{\pi} Im \frac{\Pi_o(s)}{\Pi_e(s)^2 s-\Pi_o(s)^2 M^2}
\end{equation}
In this form the propagator is a superposition of free
 propagators  where the mass squared has been replaces by $s$.
 We can therefore essentially do all calculations with propagators that have
 the free form, keeping in mind that we should integrate the end result over
 $s$ with a weight function $\rho_1(s) $ or $\rho_2(s)$ (Fig.2).

Going back to the calculation we can use the natural split of the propagator 
eq.(\ref{prop}) to P-even and P-odd parts to rewrite $\Gamma_a(p,q)_{\mu}=\Gamma_a(p,q)^{odd}+
\Gamma_a(p,q)^{even}$. As we are 
 interested in the anomalous magnetic moment we shall look at 
the P-odd part of ${\Gamma }_a(p)_{\mu}$.
 Using $(p \!\!\! / -m)u(p)=\bar u(p-q) 
(p\!\!\! / -q \!\!\! / -m)=0$ we get after some algebra :
\begin{equation}
\Gamma_a^{odd}=e^2 M \epsilon_{\lambda \nu \sigma}\int_0^{\infty} ds \rho_2(s)
\int \frac{d^3 k}{(2\pi)^3} k_{\sigma} \frac{\bar u(p-q) [2(p-q)_{\nu}+ k 
\!\!\! / 
\gamma_{\nu}]\gamma_{\mu}[2p_\lambda +\gamma_{\lambda} k \!\!\! / ]u(p)}{k^2
[k^2-s][(p-k)^2-m^2][(p-k-q)^2-m^2]} 
\end{equation} 
from the $\gamma$ matrix identities eq.(\ref{gid}) We can write $\Gamma_a^{odd}=
\Gamma_a^{(1)}+\Gamma_a^{(2)}$ Where
\begin{eqnarray}
\Gamma_a^{(1)}&=&2e^2 M \epsilon_{\lambda \nu \sigma} \int_0^{\infty} 
\frac{\rho_2(s)}{s}\Gamma_a^{(1)}(s)      \\ \nonumber
\Gamma_a^{(1)}(s)&=& \frac{1}{\sqrt{s}}
\int k_{\sigma}\frac{p_{\lambda}k \!\!\! / \gamma_{\nu}\gamma_{\mu}+
(p-q)_{\nu}\gamma_{\mu} \gamma_{\lambda}k \!\!\! /}{[(p-k-q)^2-m^2][
(p-k)^2-m^2]}\left[ \frac{1}{k^2}-\frac{1}{k^2-s}\right]  \\ \nonumber
\Gamma_a^{(2)}&=&e^2 M \int_0^{\infty} \rho_2(s)\Gamma_a^{(2)}(s)
\\ \nonumber    
\Gamma_a^{(2)}(S)&=&  \intt \frac{\epsilon^{\sigma \lambda \mu}
\gamma^{\lambda} k^{\sigma}+k_{\mu}+
 k \!\!\! / \gamma_{\mu}}{[(p-k-q)^2-m^2][
(p-k)^2-m^2][k^2-s]}.     
\end{eqnarray}
using standard Feynman parametrization we can perform the integration 
over $k$. To simplify the calculation we also assume that $ q \ll M, m$
and use the on shell condition $p^2=m^2$. Isolating the term that is 
proportional to $\epsilon_{\mu \nu \rho} p^{\mu} q^{\rho}$ we get 
\bqr
 \mu_a&=&-\frac{e^2M}{m^5} \int_o^{\infty} ds \, \frac{\rho_2(s)}{s}\left[ 6m\sqrt{s}
+8m^2 +(3s+6\sqrt{s}m)\log \left[\frac{2m\sqrt{s}-s}{4m^2-s}\right] \right]
\frac{\sqrt{s}}{\sqrt{s}+2m}  \nonumber \\
&=& \frac{1}{m \kappa}
\eqr 
we have calculated this integral numerically. In the limit $\kappa \rightarrow 
\infty$ the corrections to the value obtained in weak coupling i.e 
$\mu =\frac{1}{m}(\frac{1}{2}+\frac{1}{\kappa}) $
are down by power of $\frac{1}{\kappa}$. Thus for large $\kappa$ there is no
 difference  between the weak coupling result and the large N result even
 though the later uses an improved propagator for the gauge field.

{\it Conclusions}

In this paper we have shown that for $P$ and $T$ violating  Q.E.D in the 
anyonic limit the gyromagnetic ratio of anyons is exactly 2 to leading 
order in $1/N$. By using the large $N$ method we have furnished a strong
non-perturbative support to the result claimed in \cite{poly}.

{\it Acknowledgements}

We thank V.P. Nair for suggesting this problem, A. Kovner for a useful 
discussion and J. Simon for helping us with the graphics.
This work is partially supported by the National Sience Foundation.

{\bf Figure Captions}

Fig.1 The diagrams contributing to the anomalous magnetic moment in the
leading order in 1/N.

Fig.2 $\rho_1(p)$ and $\rho_2(p)$ for $M=4$, $\frac{e^2}{4 \pi}=0.3$ and
$m=1$
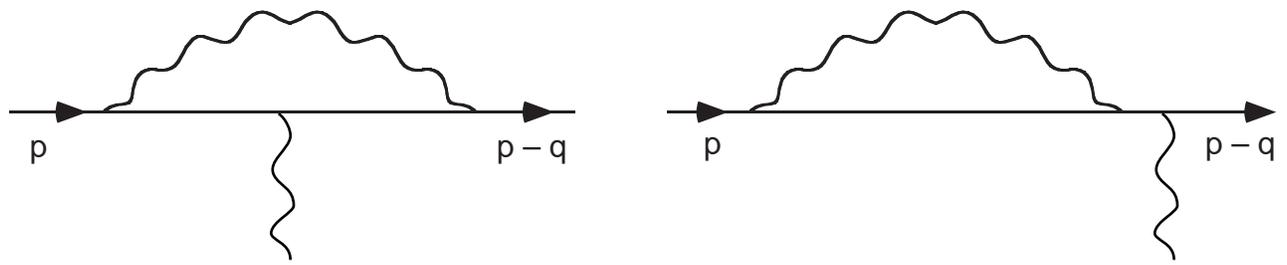
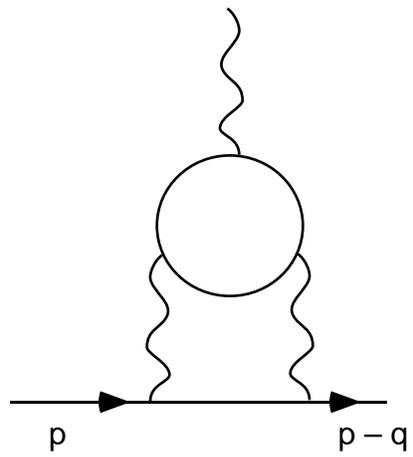
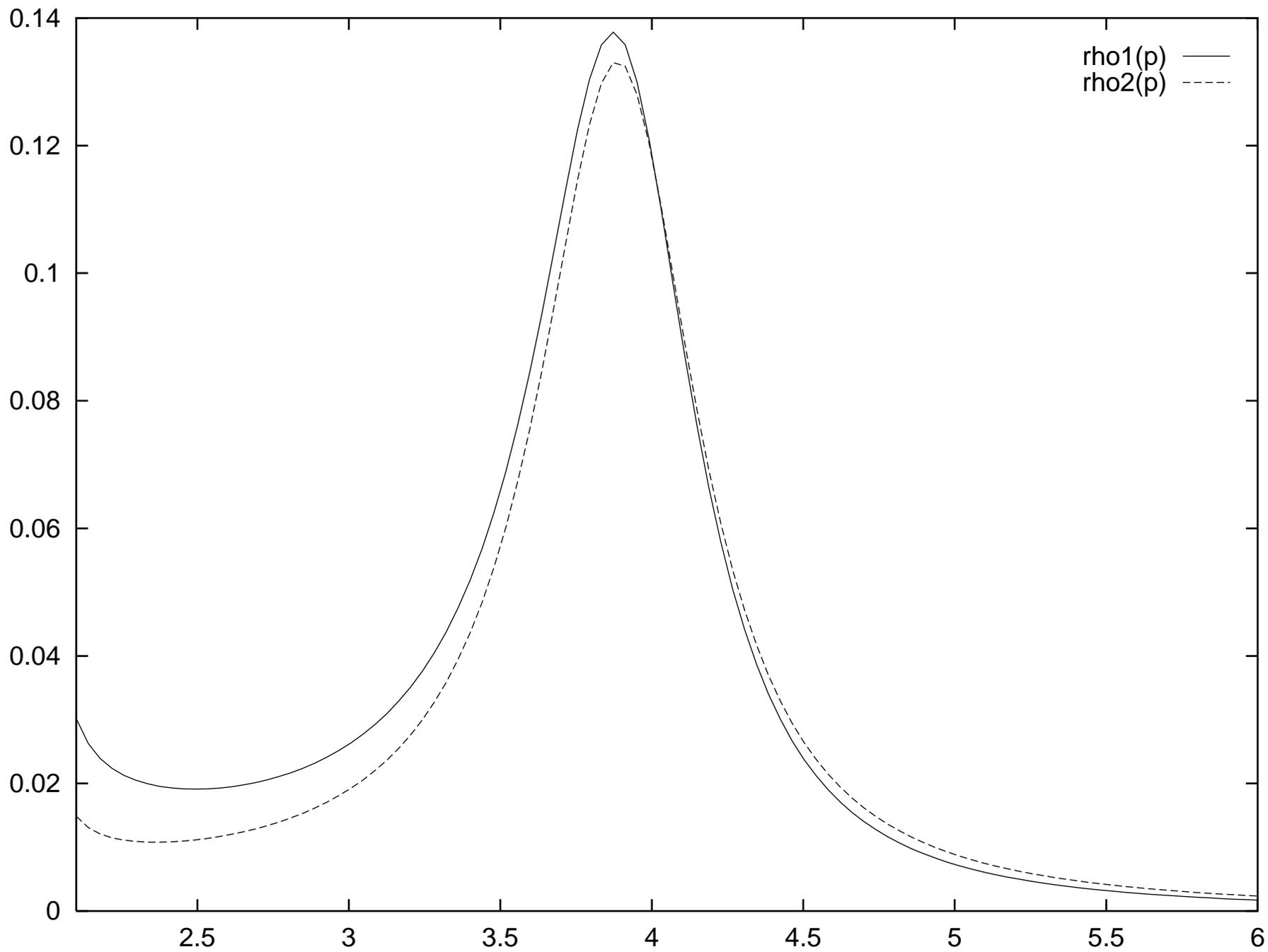
\end{document}